\documentclass[conference]{IEEEtran}
\pdfoutput=1
\IEEEoverridecommandlockouts
\usepackage{array}
\usepackage{color}
\usepackage{epsf}
\usepackage{times}
\usepackage{epsfig}
\usepackage{graphicx}
\usepackage{epstopdf}
\usepackage{cite}
\usepackage{amsmath}
\usepackage{amssymb}
\usepackage{amsfonts}
\usepackage{amsxtra}
\usepackage{amsthm}
\usepackage{algorithmic}
\usepackage{algorithm}
\usepackage{subfigure}
\usepackage{bbm}
\usepackage{textcomp}
\usepackage{xcolor}

\def\BibTeX{{\rm B\kern-.05em{\sc i\kern-.025em b}\kern-.08em
		T\kern-.1667em\lower.7ex\hbox{E}\kern-.125emX}}
\newtheorem{theorem}{Theorem}
\newtheorem{lemma}{Lemma}

\begin{document}
	
	\title{Stochastic Geometry Analysis of Hybrid Aerial Terrestrial Networks with mmWave Backhauling}

\author{\IEEEauthorblockN{Nour Kouzayha\IEEEauthorrefmark{1}, Hesham ElSawy\IEEEauthorrefmark{2}, Hayssam Dahrouj\IEEEauthorrefmark{3}, Khlood Alshaikh, Tareq Y. Al-Naffouri\IEEEauthorrefmark{4},\\ and Mohamed-Slim Alouini\IEEEauthorrefmark{4}}
	\IEEEauthorblockA{\IEEEauthorrefmark{1}Computer and Communications Engineering Department, Lebanese International University}
	\IEEEauthorblockA{\IEEEauthorrefmark{2}Electrical Engineering Department, King Fahd University of Petroleum and Minerals}
	\IEEEauthorblockA{\IEEEauthorrefmark{3}Department of Electrical Engineering, Effat University}
	\IEEEauthorblockA{\IEEEauthorrefmark{4}CEMSE, King Abdullah University of Science and Technology}}	

	\maketitle
	
\begin{abstract}
To meet increasing data demands, service providers are considering the use of Unmanned aerial vehicles (UAVs) for delivering connectivity as aerial base stations (BSs). UAVs are especially important to provide connectivity in case of disasters and accidents which may cripple completely the existing terrestrial networks. However, in order to maintain the communication of UAVs with the core network, it is essential to provide them with wireless backhaul connection to terrestrial BSs. In this work, we use stochastic geometry to study the impact of millimeter-wave (mmWave) backhauling of UAVs in a hybrid aerial-terrestrial cellular network where the UAVs are added to assist terrestrial BSs in delivering reliable service to users (UEs). In the proposed model, the UE can associate to either a terrestrial BS or a UAV connected to a BS to get backhaul support. The performance of the proposed model is evaluated in terms of coverage probability and validated against intensive simulations. The obtained results unveil that the quality of the UAVs' mmWave backhaul link has a significant impact on the UE 's experience and the deployment of UAVs must be adjusted accordingly.
\end{abstract}

	\IEEEpeerreviewmaketitle
\vspace{-0.5cm}
\section{Introduction}
Unmanned Aerial Vehicles (UAVs) are emerging as a promising solution to enhance wireless coverage in commercial cellular networks. Due to their mobility capabilities, UAVs are of particular importance in events of terrestrial cellular systems dilapidation, infrastructure absence in remote and suburban areas, and occasional events wherein there is a temporary need for supplementary network resources~\cite{Mozaffari}. In view of the UAVs’ potential as supporting solution for wireless communications, significant research efforts have been recently devoted to model, analyze and design UAVs networks. These research works have focused mainly on developing new models for air-to-ground and air-to-air channels~\cite{Khawaja}, optimizing the deployment of UAVs in terms of efficient resource allocation~\cite{Ono} and trajectory planning~\cite{Zeng} and evaluating the performance of UAV-assisted cellular networks~\cite{Wang, Lyu}. In~\cite{Wang}, the authors considered a probabilistic line-of-sight (LOS) model to evaluate the coverage probability and the area spectral efficiency of a UAV-assisted cellular network and to study the impact of different system parameters. 
\begin{figure}
	\begin{center}
		\noindent
		\includegraphics[width=0.9\linewidth]{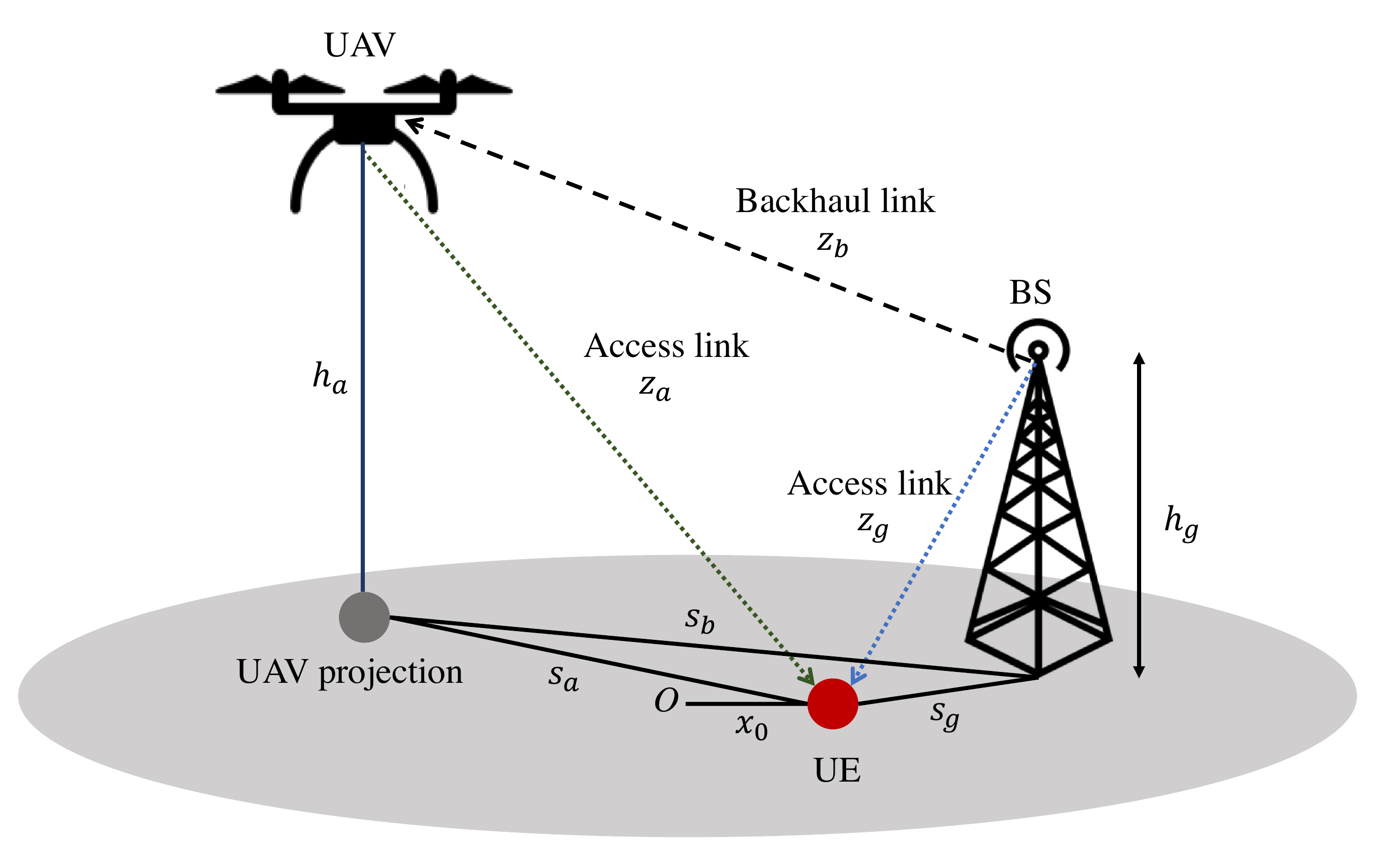}
		\caption{Proposed system model.}\label{fig:system_model}
	\end{center}
	\vspace{-1cm}
\end{figure}
Whereas terrestrial BSs connect to core network via wired/wireless backhaul, UAVs need exclusively a wireless backhaul link to connect to the core network. This link can be established through a terrestrial BS using sub-6 GHz technologies, millimeter waves (mmWave)~\cite{Xiao} or free space optical channels~\cite{Alzenad}. While the sub-6 GHz technologies allow for simpler BS deployment, mmWave backhauling provides for a superior performance due to larger bandwidths and active beam steering. In~\cite{Xiao}, the authors investigate mmWave as an enabling technology for UAV networks. The work addresses the issues of beam tracking, LOS blockage and UAV discovery. Although, the UAV backhaul has a direct impact on the ability of the UAV to provide service to UEs, the backhaul link has not been tackled effectively by the wireless community. Most of the existing works assume a guaranteed wireless backhaul link for the UAV and limit the scope of their work to the access link. An example of the papers that addressed the UAV backhaul explicitly is~\cite{Boris}, where the authors use stochastic geometry to model a UAV-based network. In this work, UAVs are placed at a fixed height to provide coverage for ground UEs while using dedicated terrestrial BSs for wireless backhaul. However, this work assumes that the UE can only associate to a UAV and the network of UAVs is modeled as a Poisson point process (PPP). Furthermore, the analytical results are provided for the special case when all the UAVs have a backhaul and the impact of the UAV backhaul is only presented in the simulation results. In this paper, we develop a general mathematical framework, based on stochastic geometry, to study the impact of the UAV backhaul in a hybrid aerial-terrestrial network, in which UAVs are introduced to assist terrestrial BSs to provide coverage for ground UEs. The introduced UAVs are backhauled with mmWave links with the terrestrial BSs to provide service to the ground UEs. To the best of the authors' knowledge, this is the first work that jointly account for the access link and mmWave backhaul link in a hybrid aerial/terrestrial cellular network where the UE can associate to either a BS or a UAV to get coverage. In addition, the UAV network is modeled as a binomial point process (BPP) which provides more realistic results in modeling a given number of UAVs in a finite region.

\vspace{-0.3cm}
\section{System Model}\label{sec:system_model}
\subsection{Network Model}
Consider the downlink (DL) of a one-tier cellular network, served by ground BSs covering a certain geographical area, and populated with UAVs providing wireless coverage for the UEs located within the area. The BSs provide wireless access to the UEs, and are connected to the UAVs through mmWave backhaul links. An example of the network model is presented in Fig.~\ref{fig:system_model}. Since we focus on a DL scenario, the UAVs transmit to UEs and receive from BSs for backhauling purposes. The terrestrial BSs are independently distributed according to a homogeneous Poisson point process (HPPP) $\Phi_g$ of density $\lambda_g$ and are all at the same height $h_g$ from the ground level. Each BS transmit power is denoted by $P_{t,g}$, which is held fixed in the context of this paper. To assist the terrestrial BSs, $N$ UAVs are distributed uniformly in a finite disk $D_c$ with radius $r_c$ forming a binomial point process (BPP) $\Phi_a$. All the UAVs hover at the same altitude $h_a$ and transmit with the same power $P_{t,a}$. Both BSs and UAVs are equipped with two antennas, one to communicate with the UE on the ground and the other for the mmWave backhaul connection. The analysis is performed at a typical single-antenna UE located at an arbitrary distance $x_0$ from the origin. 
\vspace{-0.3cm}
\subsection{Channel Model}
\subsubsection{BS-UE Access Channel}
The fading channel of the BS-UE access link consists of a large-scale fading modeled using a distance-dependent path-loss with path-loss exponent $\eta_g$, and a small-scale Rayleigh fading $\Omega_g$ with exponential distribution and unit mean. The signal power received at the UE from the $i$-th BS can thus be expressed as $P_{r,g,i}=P_{t,g}(s_{g,i}^2+h_g^2)^{-\eta_g/2}\Omega_{g,i}$, where $s_{g,i}$ is the horizontal distance separating the UE and the projection of the $i$-th BS on the ground.
\subsubsection{UAV-UE Access Channel}
The UAV-UE fading channel is characterized using the combination of two components: (a) a large-scale fading modeled using a distance-dependent path-loss with path-loss exponent $\eta_{a}$ and (b) a small scale Nakagami-$m$ fading denoted $\Omega_{a}$ and modeled using a gamma-distributed random variable with shape parameter $m_a$ and scale parameter $1/m_a$. The power received at the reference UE from the $i$-th UAV is  $P_{r,a,i}=P_{t,a}z_{a,i}^{-\eta_{a}}\Omega_{a,i}$, where $P_{t,a}$ is the UAV transmit power and $z_{a,i}$ is the distance separating the reference UE from the $i$-th UAV.
\subsubsection{BS-UAV mmWave Backhaul Channel}\label{sec:BS-UAV_model}
UAVs associate with ground BSs through mmWave wireless backhaul links. For the mmWave backhaul links, the buildings in the environment create obstacles which breaks the LOS links. The LOS probability, denoted as $P_{\mathrm{LOS}}$, depends on the environment setup and is approximated in~\cite{Hourani} as:
\begin{equation}
P_{\mathrm{LOS}}(s_{b,j,i})=\frac{1}{1+a \exp (-b[\theta(s_{b,j,i})-a])}
\label{eq:LOS_probability}
\end{equation}
where $\theta (s_{b,j,i})= \frac{180}{\pi}\arctan(|h_a-h_g|/s_{b,j,i})$ is the elevation angle, $s_{b,j,i}$ is the horizontal distance separating the projections of the $i$-th BS and $j$-th UAV on the ground plane, and $a$ and $b$ are constant values that define the environment. 

Analog beamforming is introduced at the BSs and the UAVs on the backhaul connection. Thus, we assume channel knowledge for the UAV and its associated BS so that they can steer their antennas to maximize the directionality gain. We approximate the array patterns of the UAVs and the BSs antennas on the mmWave backhaul link by the model presented in~\cite{Andrews}. For the interfering BSs, the steering angles are distributed uniformly and the gain $G_b$ of directivity for the UAVs and the BSs beamforming is a discrete random variable which follows a probability distribution as $G_k \in \{G_g G_a, G_g g_a, g_g G_a, g_g g_a\}$ with probability $p_k \in \{c_g c_a, c_g (1-c_a), (1-c_g)c_a, (1-c_g)(1-c_a)\}$, where $c_g=\frac{\theta_g}{2\pi}$, and $c_a=\frac{\theta_a}{2\pi}$. $G_s$, $g_s$, and $\theta_s$ are the gains of the main and side lobes, and the beamwidth for the BSs and UAVs ($s\in\{\mathrm{a},\mathrm{g}\}$), respectively. For the desired backhaul signal link, the directivity gain is $G_0 = G_\mathrm{g}G_\mathrm{a}$.

The fading channel between the UAV and the BS is characterized with a large-scale fading model using a distance-dependent path loss and a small scale Nakagami-$m$ fading modeled using a gamma-distributed random variable. We consider different path loss exponents and fading parameters for the LOS and the NLOS links ($m_L$ for LOS and $m_N$ for NLOS). The received power from the $i$-th BS at the $j$-th UAV is given as $P_{r,b,j,i}=P_{t,b}G_{b}C_{t}z_{b,j,i}^{-\eta_{t}}\Omega_{t,j,i}$, where $P_{t,b}$ is the BS transmit power on the Backhaul link, $\Omega_{t,j,i}$ is the small scale fading, $C_{t}$ is the path loss intercept and $\eta_{t}$ is the pathloss exponent, where $t\in\{L,N\}$ indicates if the $i$-th BS has a LOS with the $j$-th UAV.
\subsection{Association Strategy and Performance Metrics}
We aim to study the performance of a hybrid aerial-terrestrial DL cellular network where both BSs and UAVs are used to provide coverage for UEs. Furthermore, the UAVs are connected to the BSs through backhaul links, so as to extend the BSs coverage in cases of weak BS-to-UE channel gains. For the access link, we assume that the UE connects to the BS or the UAV that offers the maximum average received power. The main performance metric that we use is the DL coverage probability defined as the probability that the signal-to-interference ratio (SIR) of reference UE is higher than a predefined threshold $\beta$. The SIR is defined as $\mathrm{SIR}=\frac{P_{r}}{I_{agg}}$, where $P_{r}$ is the received signal power and $I_{agg}$ is the received aggregate interference. To be specific, $P_{r}$ and $I_{agg}$ are equivalent to $P_{r,g}$ and $I_{agg,g}$ if the UE connects to a BS, and $P_{r,a}$ and $I_{agg,a}$ if the UE connects to a UAV. For the backhaul connection, we use the minimum path loss association rule. Thus, each UAV selects the BS with the minimum average power loss to get backhaul support; the UAV then steers its antenna to align with the chosen BS. If the signal-to-interference-and-noise-ratio ($\mathrm{SINR}$) of the backhaul connection does not meet a certain threshold $\tau_b$ then the UAV is considered in outage and cannot serve the UE; otherwise the UAV can serve the UE using its end UE antenna. We define the backhaul probability as the probability that the received $\mathrm{SINR}$ of an arbitrary UAV exceeds $\tau_b$.
\vspace{-0.2cm}
\section{Analytical Results}\label{sec:unaware}
\subsection{Association Probabilities}
To get coverage, the UE can associate to either a terrestrial BS or a UAV. The UAVs are connected to the core network through mmWave backhaul links with the terrestrial BSs. We start by deriving the association probability defined as the probability that the reference UE is associated to either a ground BS or a UAV. The association probability results are provided in the lemma below.
\begin{lemma}
	\emph{Denoting by $A_g$ the probability that the reference UE at $x_0$ being served by a ground BS, this association probability is given as follows
		\begin{equation}\small
		A_{g}=2\pi\lambda_g\int_{0}^{E_{a}(w_{p})}r \exp\left(-\pi\lambda_g r^2\right)\left(\int_{E_{g}(r)}^{w_{p}}f_{W}(w)\mathrm{d}w\right)^N\mathrm{d}r
		\end{equation}
		where
		\begin{equation}
		E_{a}(x)=\sqrt{\left(\frac{P_{t,g}}{P_{t,a}}\right)^{\frac{2}{\eta_g}}x^{\frac{2\eta_{a}}{\eta_g}}-h_{g}^{2}}
		\end{equation}
		\begin{equation}
		E_{g}(x)=\left(\frac{P_{t,a}}{P_{t,g}}\right)^{\frac{1}{\eta_{a}}}\left(x^2+h_{g}^2\right)^{\frac{\eta_g}{2\eta_{a}}}
		\end{equation}
		and $f_{W}(w)$ is given in Lemma 2 of~\cite{Chetlur} as
		\begin{equation}\footnotesize
		f_{W}(w)=\begin{cases}
		f_{W_{1}}(w)=\frac{2w}{r_{c}^2}, & h_a\leq w\leq w_m\\
		f_{W_{2}}(w)=\frac{2w}{\pi r_{c}^2}\arccos\left(\frac{w^2+x_{0}^2-d^2}{2x_0\sqrt{w^2-h_{a}^2}}\right), & w_m\leq w \leq w_p
		\end{cases}
		\end{equation}
		where $r_c$ is the radius of the disk containing the set of UAVs, $w_m=\sqrt{(r_c-x_0)^2+h_a^2}$ and $w_p=\sqrt{(r_c+x_0)^2+h_a^2}$. The probability that the reference UE is connected to a UAV is $A_{a}=1-A_{g}$.}
	\begin{IEEEproof}
		\emph{See Appendix~\ref{app:association}.}
	\end{IEEEproof}
\end{lemma}
Note here that the association rule adopted by the UE creates an exclusion region on the locations of all BSs when the UE connects to a UAV. Specifically, the nearest BS must be further than $E_{a}(x_a)$ where $x_a$ is the distance to the serving UAV and $E_{a}(x)$ is given in (3). Thus, all the remaining BSs must be further that $E_{a}(x_a)$. Similarly, an exclusion region $E_g(x_g)$ is created on the locations of all the UAVs when the UE associates with a BS where $x_g$ is the distance to the serving BS and $E_g(x_g)$ is given in (4).
\subsection{UE-BS Conditional Coverage Probability}
The conditional coverage probability is defined as the probability that the received SIR is higher than the threshold $\beta$ given the UE association status. When the UE associates to a BS, the conditional coverage probability can be expressed as $P_{cov,g}=\mathbb{P}\left[\mathrm{SIR}\geq \beta | \mathrm{BS}\right]$. In the proposed model, we assume that no frequency reuse is used and the set of BSs and UAVs are sharing the same frequency resources. Thus, when the UE associates to a BS, the aggregate interference $I_{agg,g}$  includes the interference from all BSs excluding the serving BS denoted as $\hat{I}_{g}$ and the interference from all UAVs denoted as $I_{a}$. The conditional coverage probability $P_{cov,g}$ given that the UE associates to a terrestrial BS is given in the lemma below.
\begin{lemma}
	\emph{The conditional coverage probability $P_{cov,g}$ given that the UE is connected to a BS is
		\begin{equation}
		P_{cov,g}=\int_{0}^{E_a(w_p)}\mathcal{L}_{\hat{I}_g}(s_1)\mathcal{L}_{I_a}(s_1)f_{X_g}(x_g)\mathrm{d}x_{g}.
		\end{equation}
		where $s_{1}=\frac{\beta(x_g^2+h_g^2)^{\frac{\eta_g}{2}}}{P_{t,g}}$, $\mathcal{L}_{\hat{I}_g}(s_{1})$ and $\mathcal{L}_{I_a}(s_{1})$ are the Laplace transforms of the aggregate interference of all the BSs except the serving BS and of all the UAVs. $f_{X_g}(x_g)$ is the probability density function (PDF) of the conditional distance $X_g$ from the reference UE to the serving BS.}
	\begin{IEEEproof}
		\emph{See Appendix~\ref{app:pcovg}.}
	\end{IEEEproof}
\end{lemma}
To obtain the final expression of the conditional coverage probability, the Laplace transforms of the aggregate interference terms and the PDF of the conditional distance to serving BS must be computed. Lemma~$3$ and Lemma $4$ present these results as follows.
\begin{lemma}
	\emph{The Laplace transform of the interference $\hat{I}_g$ of all BSs except the serving BS to which the UE associates is
		\begin{equation}\footnotesize
		\mathcal{L}_{\hat{I}_{g}}(s_{1})=\exp\left[-2\pi\lambda_g\int_{x_g}^{\infty}\left(1-\frac{1}{s_{1}P_{t,g}(z^2+x_g^2)^{-\frac{\eta_{g}}{2}}}\right)z\mathrm{d}z\right].
		\end{equation}
		The Laplace transform of the aggregate interference $I_a$ from all the UAVs when the UE associates to a BS is given as
		\begin{equation}\small
		\begin{aligned}
		\mathcal{L}_{I_{a}}(s_{1})&=\left(\frac{1}{\int_{E_{g}(x_{g})}^{w_{p}}f_{W}(w)\mathrm{d}w}\left(\int_{E_{g}(x_{g})}^{w_{p}}\left(1+\frac{s_{1}P_{t,a}u^{-\eta_{a}}}{m_{a}}\right)^{-m_{a}}\right.\right.\\
		&\times f_{W}(u)\mathrm{d}u\bigg)\bigg)^{N}
		\end{aligned}
		\end{equation}
		where $E_{g}(x_{g})$ is the minimum distance at which the UAVs are placed when the UE associates to a BS at distance $x_g$.
	}
	\begin{IEEEproof}
		\emph{See Appendix~\ref{app:laplace}}
	\end{IEEEproof}	
\end{lemma}
The PDF of the conditional distance from the UE to the serving BS $X_g$ is provided in the following lemma.
\begin{lemma}
	\emph{When the UE associates to a terrestrial BS, the PDF of the distance to the serving BS is
		\begin{equation}\small
		f_{X_{g}}(x_g)=\frac{1}{A_{g}}2\pi\lambda_g x_g \exp\left(-\pi\lambda_g x_{g}^2\right)\left(\int_{E_{g}(x_g)}^{w_p}f_{W}(w)\mathrm{d}w\right)^N
		\end{equation}
		where $A_{g}$, $E_{g}(x_g)$ and $f_W(w)$ are given in Lemma 1.
	}
	\begin{IEEEproof}
		\emph{See Appendix~\ref{app:distance}.}
	\end{IEEEproof}
\end{lemma}
\subsection{UE-UAV Conditional Coverage Probability}
When the UE chooses to associate to a UAV, this UAV must have a strong mmWave backhaul link with a terrestrial BS to be able to provide coverage for the reference UE. Thus, two conditions are required for coverage:
\begin{itemize}
\item $\mathrm{SIR}>\beta$: The probability that the received $\mathrm{SIR}$ at the UE from its serving UAV must exceed a threshold $\beta$.
\item  $\mathrm{SINR}>\tau_b$: The received $\mathrm{SINR}$ at the serving UAV from the BS to which it connects for backhaul support needs to be greater than a specific threshold $\tau_b$.
\end{itemize}
Therefore, the conditional coverage probability given that the UE associates to a UAV is the joint probability of these two events and can be expressed as $P_{cov,a}=\mathbb{P}\left[\mathrm{SIR}\geq \beta | \mathrm{UAV},\mathrm{SINR}\geq\tau_b\right]$, where the first term corresponds to the coverage condition and the second term is the added backhaul link quality constraint. We assume that these two events are independent so that $P_{cov,a}=\mathbb{P}\left[\mathrm{SIR}\geq \beta | \mathrm{UAV}\right]\times\mathbb{P}\left[\mathrm{SINR}\geq\tau_b\right]$ and we prove the accuracy of such assumption by comparing the analytical results with the simulations in Section~\ref{sec:results}. 

We start by providing an expression for the backhaul probability $S(\tau_b)=\mathbb{P}\left[\mathrm{SINR}\geq\tau_b\right]$. Without loss of generality, the analysis is performed for a reference UAV located at height $h_a$ and at the center of the disk $D_c$. According to the backhaul rule, the UAV connects to the ground BS providing the lowest path loss. Due to the distance-dependent LOS probability defined in (1), the set of BSs is divided into two sub-processes: The first one in the LOS BSs set $\Phi_L$ which includes the BSs having LOS links with the reference UAV and has a density $\lambda P_{\mathrm{LOS}}$(r), where $r$ is the distance from the reference UAV. The second one is the NLOS BSs set $\Phi_N$ with density $\lambda(1-P_{\mathrm{LOS}}(r))$. Since the UAV connects to the BS with the smallest path loss to get backhaul support, then the serving BS will be the nearest BS in $\phi_L$ or the nearest BS in $\phi_N$. The following lemma provides expressions for the probabilities that the reference UAV connects with a LOS BS or a NLOS BS.
\begin{lemma}
	\emph{The probability $A_{\mathrm{L}}$ that the reference UAV is connected to a LOS BS is given as
		\begin{equation}
		A_{\mathrm{L}}=\int_{0}^{\infty}e^{-2\pi\lambda_g\int_{0}^{E_L(x)}(1-P_{\mathrm{LOS}}(t))t\mathrm{d}t}f_{s_{L}}(x)\mathrm{d}x
		\end{equation}
		where $E_L(x)=\sqrt{\left(\frac{C_{L}}{C_{N}}\right)^{\frac{2}{\eta_{N}}}(x^2+\Delta_h^2)^{\frac{\eta_{L}}{\eta_{N}}}-\Delta_h^2}$, $\Delta_h=|h_a-h_g|$ and $f_{s_{L}}(x)=2\pi\lambda_g x P_{\mathrm{LOS}}(x)e^{-2\pi\lambda_g\int_{0}^{x}P_{\mathrm{LOS}}(r)r\mathrm{d}r}$. The probability that the reference UAV connects to a NLOS BS is $A_{\mathrm{N}}=1-A_{\mathrm{L}}$.}
	\begin{IEEEproof}
		\emph{See Appendix \ref{app:LOS_association}.}
	\end{IEEEproof}
\end{lemma}
The distances from the reference UAV to the serving BS in $\Phi_L$ and $\Phi_N$ are denoted by $X_L$ and $X_N$, respectively and are provided in the following lemma.
\begin{lemma}	
	\emph{Given that a UAV connects to a LOS BS to get backhaul support, the PDF of the distance $X_{L}$ to its serving BS is
		\begin{equation}
		f_{\mathrm{L}}(x)=\frac{f_{s_{L}}(x)}{A_{\mathrm{L}}}e^{-2\pi\lambda_g\int_{0}^{E_L(x)}(1-P_{\mathrm{LOS}}(t))t\mathrm{d}t},
		\end{equation}
		Given that the UAV associates with a NLOS BS in $\Phi_N$, the PDF of the distance $X_{N}$ to the serving BS is
		\begin{equation}
		f_{\mathrm{N}}(x)=\frac{f_{s_{N}}(x)}{A_{\mathrm{N}}}e^{-2\pi\lambda_g\int_{0}^{E_N(x)}P_{\mathrm{LOS}}(t)t\mathrm{d}t},
		\end{equation}
		where $E_N(x)=\sqrt{\left(\frac{C_{N}}{C_{L}}\right)^{\frac{2}{\eta_{L}}}(x^2+\Delta_h^2)^{\frac{\eta_{N}}{\eta_{L}}}-\Delta_h^2}$
		and $f_{s_N}(x)=2\pi\lambda_g x (1-P_{\mathrm{LOS}}(x))e^{-2\pi\lambda_g\int_{0}^{x}(1-P_{\mathrm{LOS}}(r))r\mathrm{d}r}$.}
	\begin{IEEEproof}
		\emph{See Appendix~\ref{app:backhaul_distance}.}
	\end{IEEEproof}
\end{lemma}
Finally, the overall backhaul probability is presented in Theorem~$1$.
\begin{theorem}[Backhaul Probability]
	\emph{The backhaul probability $S(\tau_b)$ can be derived as
		\begin{equation}
		S(\tau_b)=A_{\mathrm{L}}S_{\mathrm{L}}(\tau_b)+A_{\mathrm{N}}S_{\mathrm{N}}(\tau_b)
		\end{equation}
		where $S_\mathrm{L}(\tau_b)$ and $S_\mathrm{N}(\tau_b)$ are the conditional backhaul probabilities given that the reference UAV is connected to a LOS BS or a NLOS BS. $A_{\mathrm{L}}$ and $A_{\mathrm{N}}$ are the corresponding association probabilities. The conditional backhaul coverage probabilities $S_{\mathrm{L}}$ and $S_{\mathrm{N}}$ are given as follows
		\begin{equation}
		\begin{aligned}
		&S_{\mathrm{L}}(\tau_b)\approx\sum_{n=1}^{m_\mathrm{L}}(-1)^{n+1} {m_\mathrm{L}\choose n}\\
		&\int_{0}^{\infty}e^{-\frac{n \gamma_\mathrm{L}\left(x^2+\Delta_h^2\right)^{\frac{\eta_L}{2}}\tau_b\sigma^2}{P_{t,g}C_{\mathrm{L}}G_{0}}-Q_n(\tau_b,x)-V_n(\tau_b,x)}f_L(x)\mathrm{d}x
		\end{aligned}
		\end{equation}
		\begin{equation}
		\begin{aligned}
		&S_{\mathrm{N}}(\tau_b)\approx\sum_{n=1}^{m_\mathrm{N}}(-1)^{n+1} {m_\mathrm{N}\choose n}\\
		&\int_{0}^{\infty}e^{-\frac{n \gamma_\mathrm{N}\left(x^2+\Delta_h^2\right)^{\frac{\eta_N}{2}}\tau_b\sigma^2}{P_{t,g}C_{\mathrm{N}}G_{0}}-W_n(\tau_b,x)-Z_n(\tau_b,x)}f_N(x)\mathrm{d}x
		\end{aligned}
		\end{equation}
		where
		\begin{equation}\small
		\begin{aligned}
		Q_n(\tau_b,x)&=2\pi\lambda_g\sum_{k=1}^{4}p_k\int_{x}^{\infty}F\left(m_\mathrm{L},\frac{n\gamma_\mathrm{L}\bar{G}_k\tau_b(x^2+\Delta_h^2)^{\frac{\eta_L}{2}}}{m_{\mathrm{L}}(t^2+\Delta_h^2)^{\frac{\eta_L}{2}}}\right)\\
		&\times P_{\mathrm{LOS}}(t)t\mathrm{d}t
		\end{aligned}
		\end{equation}
		\begin{equation}\small
		\begin{aligned}
		V_n(\tau_b,x)&=2\pi\lambda_g\sum_{k=1}^{4}p_k\int_{E_L(x)}^{\infty}F\left(m_\mathrm{N},\frac{nC_{\mathrm{N}}\gamma_\mathrm{L}\bar{G}_k\tau_b(x^2+\Delta_h^2)^{\frac{\eta_L}{2}}}{C_\mathrm{L}m_{\mathrm{N}}(t^2+\Delta_h^2)^{\frac{\eta_N}{2}}}\right)\\
		&\times\left(1-P_{\mathrm{LOS}}(t)\right)t\mathrm{d}t
		\end{aligned}
		\end{equation}
		\begin{equation}\footnotesize
		\begin{aligned}
		W_n(\tau_b,x)&=2\pi\lambda_g\sum_{k=1}^{4}p_k\int_{E_N(x)}^{\infty}F\left(m_\mathrm{L},\frac{nC_{\mathrm{L}}\gamma_\mathrm{N}\bar{G}_k\tau_b(x^2+\Delta_h^2)^{\frac{\eta_N}{2}}}{m_{\mathrm{L}}C_{\mathrm{N}}(t^2+\Delta_h^2)^{\frac{\eta_L}{2}}}\right)\\
		&\times P_{\mathrm{LOS}}(t)t\mathrm{d}t
		\end{aligned}
		\end{equation}
		\begin{equation}\small
		\begin{aligned}
		Z_n(\tau_b,x)&=2\pi\lambda_g\sum_{k=1}^{4}p_k\int_{x}^{\infty}F\left(m_\mathrm{N},\frac{n\gamma_\mathrm{N}\bar{G}_k\tau_b(x^2+\Delta_h^2)^{\frac{\eta_N}{2}}}{m_{\mathrm{N}}(t^2+\Delta_h^2)^{\frac{\eta_N}{2}}}\right)\\
		&\left(1-P_{\mathrm{LOS}}(t)\right)t\mathrm{d}t
		\end{aligned}
		\end{equation}
		where $F(m,x)=1-1/(1+x)^m$, $\gamma_L=m_{\mathrm{L}}(m_L!)^{-\frac{1}{m_L}}$ and $\gamma_N=m_{\mathrm{N}}(m_N!)^{-\frac{1}{m_N}}$  are the Nakagami-m small scale fading parameters; for $k\in\{1,2,3,4\}$, $\bar{G}_k=\frac{G_k}{G_{0}}$, $G_k$ and $p_k$ are defined in Section~\ref{sec:BS-UAV_model}.}
	\begin{IEEEproof}
		\emph{See Appendix \ref{app:backhaul_probability}.}
	\end{IEEEproof}
\end{theorem}
Finally, the coverage probability $P_{cov,a}$ when the UE associates to a UAV is given in the lemma below.
\begin{lemma}
	\emph{The conditional coverage probability $P_{cov,a}$ given that the UE is connected to a UAV is
		\begin{equation}\footnotesize
		P_{cov,a}=S(\tau_b) \int_{h_a}^{w_p}\sum_{k=0}^{m_a-1}\frac{(-s_2)^k}{k!}\left[\frac{\partial^k}{\partial s_{2}^{k}}\mathcal{L}_{I_g}(s_2)\mathcal{L}_{\hat{I}_a}(s_2)\right] f_{X_a}(x_a)\mathrm{d}x_a.
		\end{equation}
		where $s_{2}=\frac{m_a\beta x_a^{\eta_a}}{P_{t,a}}$ and $f_{X_a}(x_a)$ is the PDF of the conditional distance to the serving UAV. $\mathcal{L}_{\hat{I}_a}(s_{2})$ and $\mathcal{L}_{I_g}(s_{2})$ are the Laplace transforms of the aggregate interference of all the UAVs except the serving UAV $\hat{I}_{a}$ and of all the BSs $I_{g}$.}
	\begin{IEEEproof}
		\emph{See Appendix~\ref{app:pcova}.}
	\end{IEEEproof}
\end{lemma}
As noted in (20), the Laplace transforms of the aggregate interference terms $\mathcal{L}_{I_g}(s_2)$ and $\mathcal{L}_{\hat{I}_a}(s_2)$ and the PDF of the conditional distance to the serving UAV $f_{X_a}(x_a)$ must be computed to obtain the final expression of $P_{cov,a}$. Lemma~$8$ presents these Laplace transforms as follows.
\begin{lemma}
	\emph{The Laplace transform of the interference $I_g$ of all BSs when the UE associates to a UAV is given as
		\begin{equation}\footnotesize
		\mathcal{L}_{I_{g}}(s_{2})=\exp\left[-2\pi\lambda_g\int_{E_{a}(x_a)}^{\infty}\left(1-\frac{1}{s_{2}P_{t,g}(z^2+x_g^2)^{-\frac{\eta_{g}}{2}}}\right)z\mathrm{d}z\right].
		\end{equation}
		where $E_{a}(x_a)$ is the minimum distance at which the BSs can be placed. The Laplace transform of the aggregate interference $\hat{I}_a$ from all the UAVs except the serving UAV is given as
		\begin{equation}
		\begin{aligned}
		\mathcal{L}_{\hat{I}_{a}}(s_{2})&=\left(\frac{1}{\int_{x_{a}}^{w_{p}}f_{W}(w)\mathrm{d}w}\left(\int_{x_{a}}^{w_{p}}\left(1+\frac{s_{2}P_{t,a}u^{-\eta_{a}}}{m_{a}}\right)^{-m_{a}}\right.\right.\\
		&\times f_{W}(u)\mathrm{d}u\bigg)\bigg)^{N-1}
		\end{aligned}
		\end{equation}
		where $f_{W}(w)$ is given in (5). }
	\begin{IEEEproof}
		\emph{The Laplace transform of the aggregate interference from all BSs $\mathcal{L}_{I_g}(s_{2})$ is obtained by following a similar approach to the proof of Lemma 3 while replacing the lower bound of the integral in (10) by $E_{a}(x_a)$. The aggregate interference from all UAVs except the serving UAV can be expressed as $\sum_{i=1}^{N-1}P_{t,a}u_{a,i}^{-\eta_{a}}\Omega_{a,i}$ and is obtained by following a similar approach to Lemma~$3$ as
			\begin{equation}\footnotesize
			\mathcal{L}_{\hat{I}_a}(s_{2})=\left(\int_{x_a}^{w_{p}}f_{U}\left(u,x_a\right)\left(1+\frac{s_{2}P_{t,a}u^{-\eta_{a}}}{m_{a}}\right)^{-m_{a}}\mathrm{d}u\right)^{N-1}
			\end{equation}
			The final expression of $\mathcal{L}_{I_a}(s_{2})$ is obtained by plugging $f_{U}(u,x_a)$ in the above equation.
		}
	\end{IEEEproof}	
\end{lemma}
Next, we derive the PDF of the conditional distance $X_a$ from the UE to the serving UAV and present it in Lemma~$9$.
\begin{lemma}
	\emph{When the UE associates to a UAV, the PDF of the distance to the serving UAV is given as
		\begin{equation}\small
		f_{X_{a}}(x_a)=\frac{N}{A_{a}}f_{W}(x_a)\exp\left(-\pi\lambda_g E_{a}(x_a)^2\right)\left(\int_{x_a}^{w_p}f_{W}(w)\mathrm{d}w\right)^{N-1}
		\end{equation}
		where $f_W(w)$, $A_a$, and $E_a(x_a)$ are given in Lemma 1.
	}
	\begin{IEEEproof}
		\emph{This proof follows a similar approach to Lemma~$4$. Thus, the PDF of $X_a$ has the following expression
			\begin{equation}
			f_{X_{a}}(x_{a})=\frac{1}{A_{a}}\bar{F}_{R_{g}}\left(E_{a}(x_{a})\right)f_{R_{a}}(x_{a})
			\end{equation}
			where $\bar{F}_{R_{g}}\left(r)\right)=\exp\left(-\pi\lambda_g r^2\right)$ is the complementary cumulative distribution function (CCDF) of the distance $R_g$ to the nearest BS and $f_{R_{a}}(x_{a})=N f_{W}(x_a)\left(\int_{x_a}^{w_p}f_{W}(w)\mathrm{d}w\right)^{N-1}$ is the PDF of the distance to the nearest UAV which can be obtained by deriving (28). Finally, by replacing $\bar{F}_{R_g}(E_a(x_a))$ and $f_{R_{a}}(x_{a})$ with their corresponding expressions, we can get the expression in (24) characterizing the distance distribution of $X_a$.
		}
	\end{IEEEproof}
\end{lemma}
\subsection{Overall Coverage Probability}
After deriving the association probabilities, the conditional distance distributions and the conditional coverage probabilities, we can get the overall coverage probability of the considered terrestrial/aerial hybrid system with mmWave backhaul capability through the total probability theorem as
\begin{equation}
	P_{cov}=A_a P_{cov,a}+A_g P_{cov,g}
\end{equation}

\vspace{-0.5cm}
\section{Numerical results and Discussions}\label{sec:results}
\begin{figure}
	\begin{center}
		\noindent
		\includegraphics[width=0.85\linewidth]{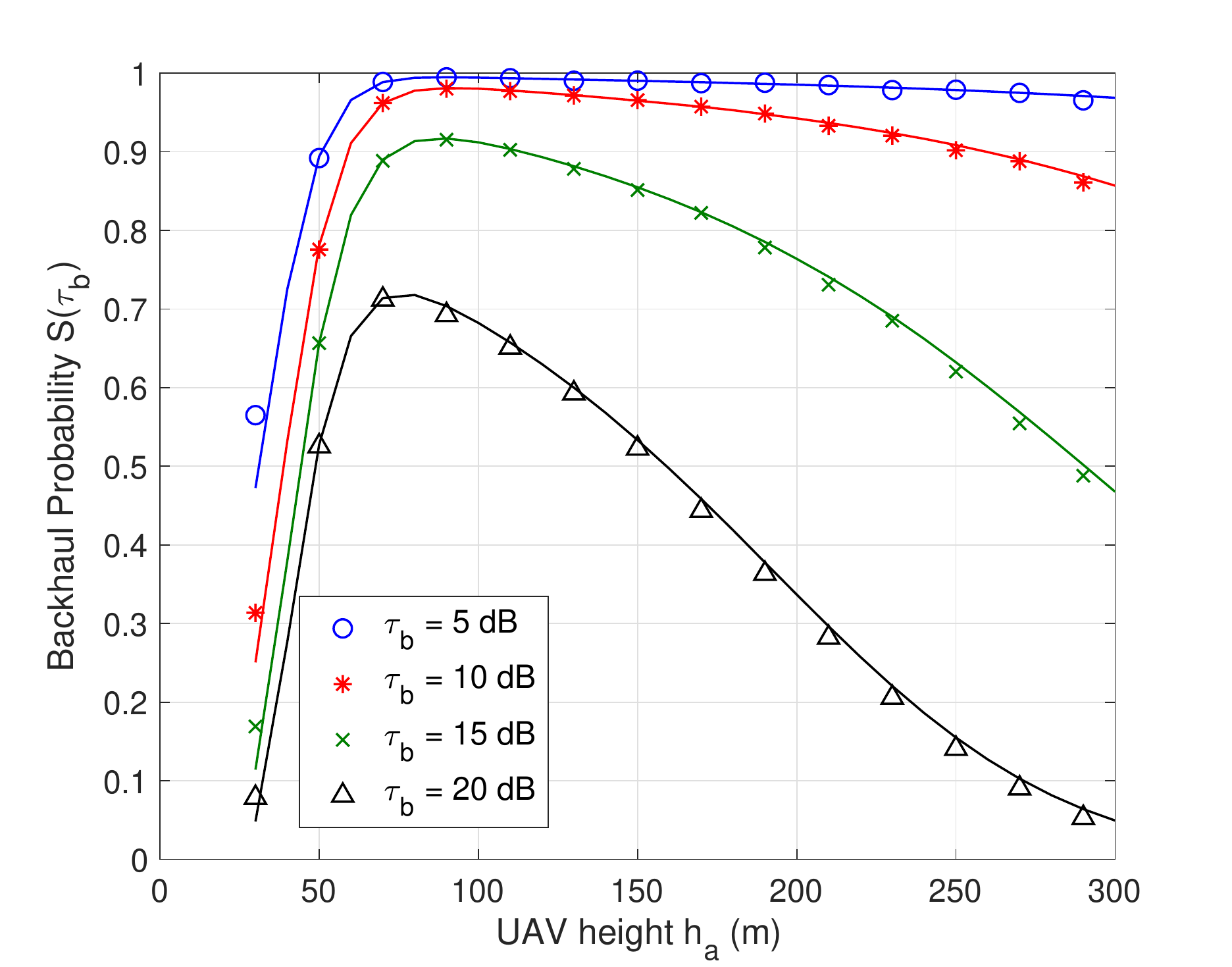}
		\caption{Backhaul probability as function of the UAV heights for different backhaul threshold values with $G_s=18$~dB, $g_s=-2$~dB and $\theta_s=20^{\circ}$, $s\in\{a,g\}$.}\label{fig:backhaul_ha}
	\end{center}
	\vspace{-0.8cm}
\end{figure}
In this section, we present performance results obtained from the proposed analytical framework and validated through extensive Monte Carlo simulations to analyze the impact of quality of the backhaul link on the coverage probability. Unless otherwise stated, we assume that the UE is at the origin ($x_0=0$) and we set the environment parameters $a=4.88$ and $b=0.43$~\cite{Hourani}. The pathloss exponents are set to $\eta_{g}=4$, $\eta_{a}=2.5$, $\eta_{L}=2.5$ and $\eta_{N}=4$, while $C_L=C_N=-69.8$~dB and the noise power is equal to $4\cdot 10^{-11}$~W~\cite{Elshaer}. The radius of the considered area is $r_{c}=1$~Km. The transmit power of the BS on the access link and the backhaul link are set to $P_{t,g}=20$~W and $P_{t,b}=10$~W, while $h_g=30$~m and $P_{t,a}=1$~W. The Nakagami-m fading parameters for the UE-UAV access link and the backhaul link are set to $m_a=3$, $m_L=3$ and $m_N=1$, while we assume Rayleigh fading with unit mean for the UE-BS access link. We start by analyzing the impact of the UAV height in Fig~\ref{fig:backhaul_ha} on the capability of the UAV to get a successful backhaul with a BS on the ground. The solid lines represent the analytical results of our model and the markers represent the simulation results. It is clearly seen in Fig.~\ref{fig:backhaul_ha} that the analytical results match perfectly with the simulations for different parameters which validates our proposed model. An optimal height for UAV deployment can also be observed to achieve the highest backhaul probability. When a UAV flies at higher height, its distance to the serving BS increases which causes a degradation in the received signal power. However, a higher altitude means a higher probability of finding a LOS link between the UAV and the BS. Thus, when flying at a low altitude, the LOS probability increases rapidly and improves the received signal quality which compensates for the power degradation. However, for high altitudes, the UAV gets a LOS link with a probability almost equal to $1$ and thus increasing its height will not improve the backhaul probability. 
\begin{figure}
	\begin{center}
		\noindent
		\includegraphics[width=0.85\linewidth]{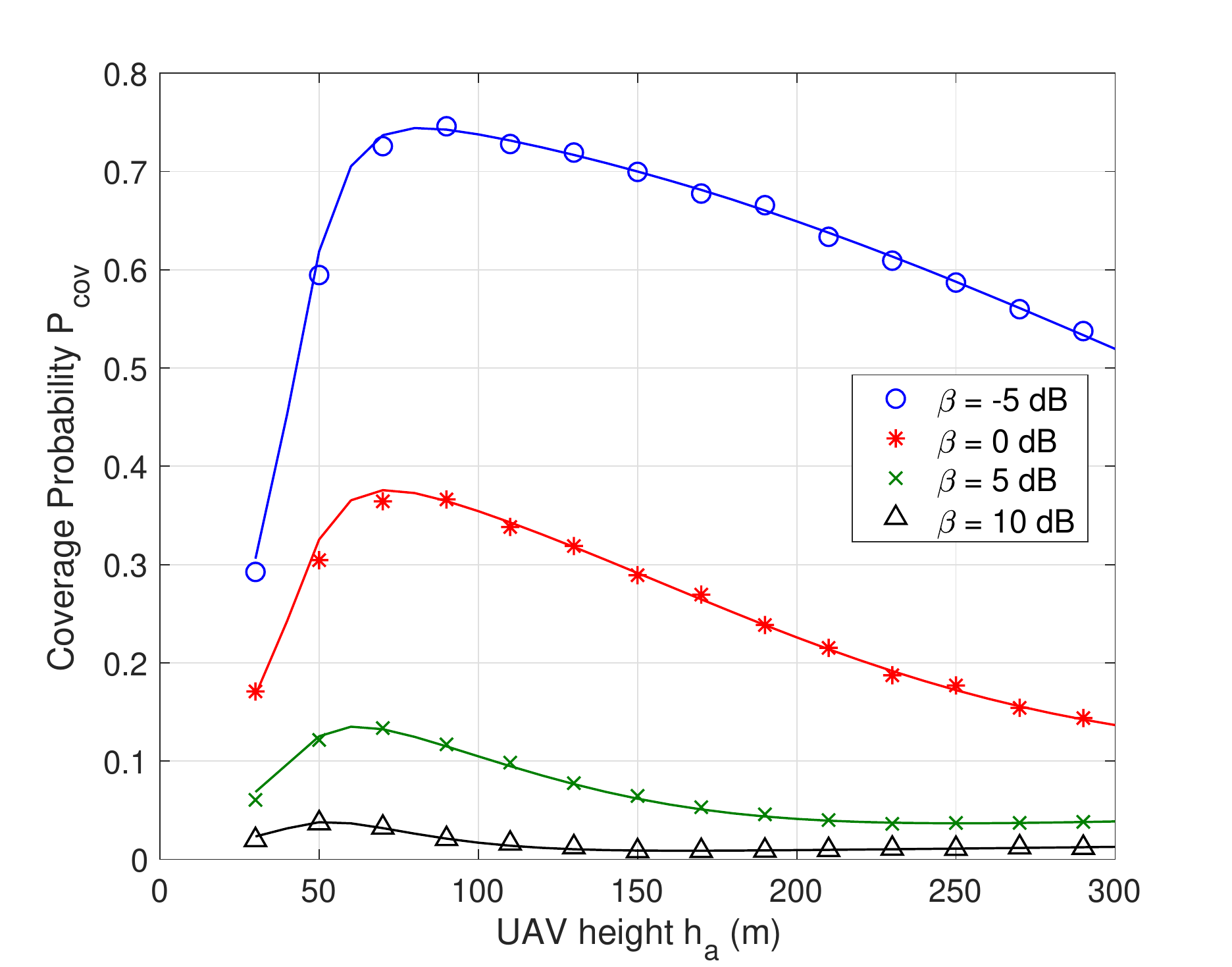}
		\caption{Coverage probability as function of the UAV heights for different coverage threshold values.}\label{fig:coverage_ha}
	\end{center}
	\vspace{-0.8cm}
\end{figure}
Fig.~\ref{fig:coverage_ha} presents the impact of the height of the UAVs on the overall coverage probability for different SIR threshold values. In the low altitudes region, the coverage probability increases as the UAV height increases. This is due to the fact that, as the height of the UAV increases, more UAVs will have successful backhaul links with  terrestrial BSs and will be able to better serve the UEs. For higher altitudes, the backhaul link quality deteriorates as shown in Fig.~\ref{fig:backhaul_ha} and the distance from the reference UE to its serving device and interfering UAVs increases, thus, the SIR and the overall coverage probability decreases. For a guaranteed UAV backhaul scenario, the coverage probability will only deteriorates with the increase of the UAV height. 
\begin{figure}
	\begin{center}
		\noindent
		\includegraphics[width=0.85\linewidth]{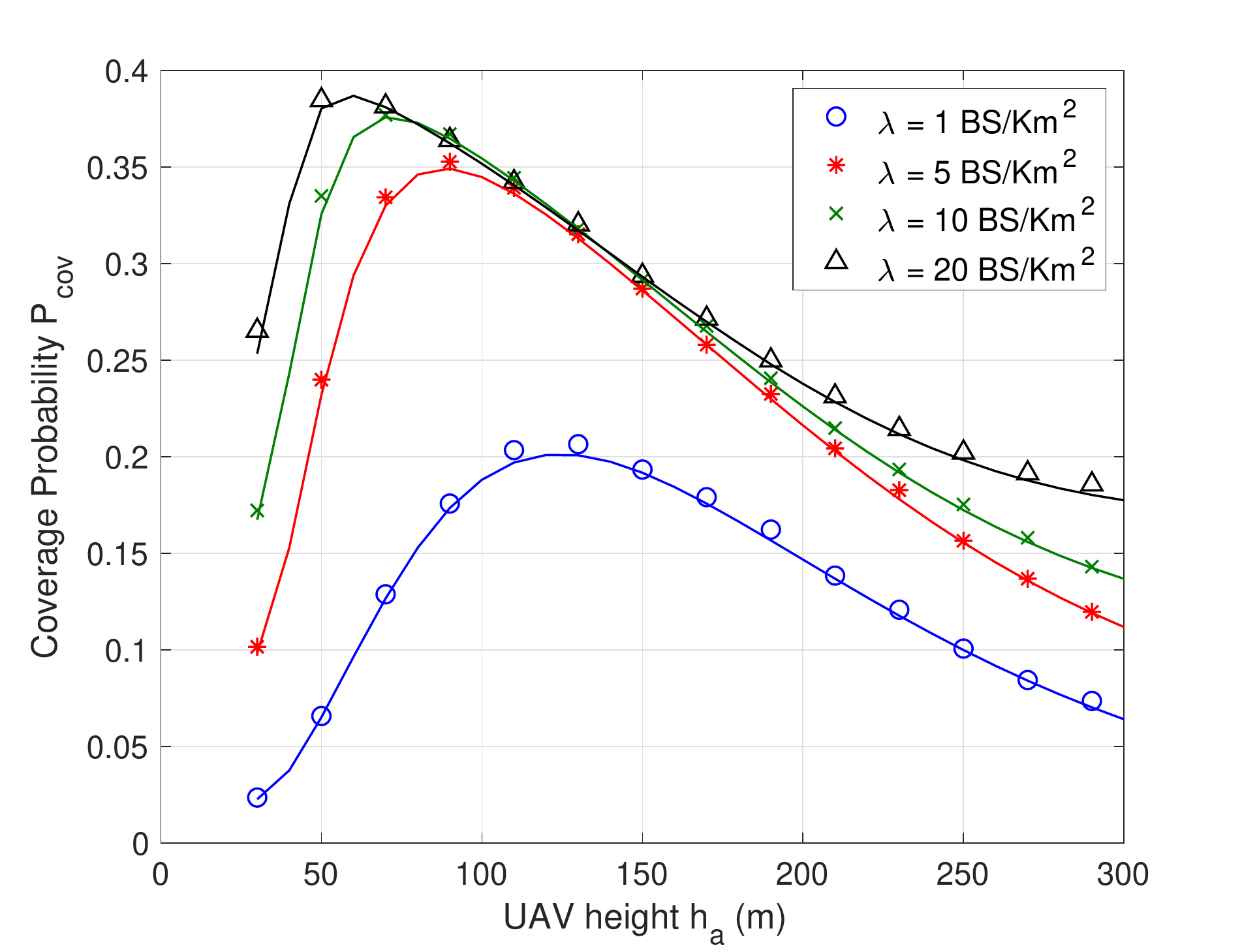}
		\caption{Coverage probability as function of the UAVs height for different BSs densities.}\label{fig:coverage_lambda}
	\end{center}
	\vspace{-0.8cm}
\end{figure}
Finally, Fig~\ref{fig:coverage_lambda} presents the coverage probability as function of the density of terrestrial BSs deployed in the network and the height of the UAVs. We can see that, for a given height, adding more BSs increases the coverage probability. Higher densities of BSs lead to a higher backhaul probability since the distance between a UAV and the BS to which it connects decreases, leading to the increase of the average received signal power from the serving BS and an increase of the LOS probability. Fig.~\ref{fig:coverage_lambda} also reveals that the increase in the density of deployed BSs adds constraints on the maximum height that the UAVs can reach while providing service to the UEs.
\vspace{-0.1cm}
\section{Conclusion}\label{sec:conclusion}
In this paper, we used stochastic geometry to assess the performance of mmWave backhauling for UAVs in a hybrid aerial-terrestrial cellular network considering key system parameters such as the UAV and the BSs heights and densities and the backhaul coverage requirement. After characterizing the backhaul probability, the association probabilities and the distance distributions, we have obtained an expression for the overall coverage probability as function of these parameters and validated our results using Monte Carlo simulations. The obtained results show that the quality of the UAVs backhaul link has a significant impact on the UE experience and adds limitations on the maximum height and number that UAVs can reach while remaining able to provide reliable service.
\vspace{-0.1cm}
\appendix
\subsection{Proof of Lemma 1}\label{app:association}
According to the association rule described in Section~\ref{sec:system_model}, the reference UE associates to the device that offers the maximum average received power. This translates to the fact that the reference UE associates to a BS if the nearest BS provides a higher received average power than the nearest UAV. Denoting by $R_a$ the distance to the nearest UAV and by $R_g$ the horizontal distance to the nearest BS, the probability $A_g$ that the UE associates to a BS can be derived as follows
\begin{equation}\small
\begin{aligned}
&A_{g}=\mathbb{P}\left[P_{t,g}\left(R_{g}^2+h_{g}^2\right)^{-\frac{\eta_g}{2}}>P_{t,a}R_{a}^{-\eta_{a}}\right]\\
&=\int_{0}^{\infty}\mathbb{P}\left[R_a>E_{g}(r)\right]f_{R_{g}}(r)\mathrm{d}r=\int_{0}^{\infty}\bar{F}_{R_{a}}(E_{g}(r))f_{R_{g}}(r)\mathrm{d}r
\end{aligned}
\end{equation}
where $E_{g}(x)$ is given in (4) and $f_{R_{g}}(r)=2\pi\lambda_g\exp\left(-\pi\lambda_g r^2\right)$ is the PDF of the horizontal distance separating the nearest BS from the reference UE~\cite{Moltchanov}. $\bar{F}_{R_{a}}(r)$ corresponds to the CCDF of the distance to the nearest UAV. Denoting by $W$ the distance from an arbitrary UAV to the reference UE, the term $\bar{F}_{R_{a}}(r)$ can be obtained following the proof of Lemma 3 in~\cite{Chetlur} as
\begin{equation}
\bar{F}_{R_{a}}(x)=\left(1-F_{W}(x)\right)^N=\left(\int_{x}^{w_p}f_{W}(w)\mathrm{d}w\right)^{N}
\end{equation}
where $F_{W}(x)$ is the cumulative distribution function (CDF) of $W$ and its PDF $f_W(w)$ is given in (5)~\cite{Chetlur}. Finally, the association probability $A_g$ can be obtained by plugging (28) in (27) and integrating it over the feasible region of $R_g$. Note that the largest value that $R_g$ can take in this case is not $\infty$. This is clearly noted in the upper limit of the outer integral in (2) being equal to $E_a(r)$, where $E_a(r)$ is given in (3). Since $w_p$ is the farthest distance between any UAV and the reference UE, the signal received from the nearest BS is lower than that of any UAV if this BS is located further than $E_a(w_p)$.
\subsection{Proof of Lemma 2}\label{app:pcovg}
The conditional coverage probability $P_{cov,g}$ is calculated as
\begin{equation}\footnotesize
\begin{aligned}
&P_{cov,g}=\mathbb{P}\left[\mathrm{SIR}\geq \beta | \mathrm{BS}\right]=\mathbb{P}\left[\frac{P_{t,g}(x_{g}^2+h_{g}^2)^{-\eta_g/2}\Omega_{g}}{I_{agg,g}}\geq \beta | \mathrm{BS}\right]\\
&=\mathbb{E}_{X_g}\left[\mathbb{E}_{I_{agg,g}}\left[\mathbb{P}\left(\Omega_{g}\geq\frac{\beta I_{agg,g}}{P_{t,g}(x_{g}^2+h_{g}^2)^{-\eta_g/2}}\right)\right]\right]\\
&\stackrel{(a)}{=}\mathbb{E}_{X_g}\left[\mathbb{E}_{I_{agg,g}}\left[\exp\left(-\frac{\beta (\hat{I}_g+I_a)}{P_{t,g}(x_{g}^2+h_{g}^2)^{-\eta_g/2}}\right)\right]\right]\\
&\stackrel{(b)}{=}\int_{0}^{E_a(w_p)}\mathcal{L}_{\hat{I}_g}(s_1)\mathcal{L}_{I_a}(s_1)f_{X_g}(x_g)\mathrm{d}x_{g}.
\end{aligned}
\end{equation}
where $s_1=\frac{\beta(x_g^2+h_g^2)^{\frac{\eta_g}{2}}}{P_{t,g}}$, (a) follows from the exponential distribution of the UE-BS small scale fading gain $\Omega_{g}$ and from the expression of the interference $I_{agg,g}=\hat{I}_{g}+I_a$. (b) follows from the independence of $\hat{I}_g$ and $I_{a}$ and from the definition of the Laplace transform that $\mathbb{E}_{I}\left[\exp(-sI)\right]=\mathcal{L}_{I}(s)$.
\subsection{Proof of Lemma 3}\label{app:laplace}
By definition, the aggregate interference from the BSs except the serving BS denoted as $b_0$ can be expressed as $\hat{I}_{g}=\sum_{s_{g,i}\in \phi_g \backslash b_0}P_{t,g}(s_{g,i}^2+h_{g}^2)^{-\eta_{g}/2}\Omega_{g,i}$. Thus, $\mathcal{L}_{\hat{I}_g}(s_{1})$ can be derived as
\begin{equation}\footnotesize
	\begin{aligned}
	\mathcal{L}_{\hat{I}_g}(s_{1})&=\mathbb{E}_{\phi_g}\left[e^{-s_{1}\sum_{s_{g,i}\in \phi_g \backslash b_0}P_{t,g}(s_{g,i}^2+h_{g}^2)^{-\frac{\eta_{g}}{2}}\Omega_{g,i}}\right]\\
	&\stackrel{(a)}{=}\mathbb{E}_{\phi_g}\left[\prod_{s_{g,i}\in \phi_g \backslash b_0}^{}\mathbb{E}_{\Omega_{g,i}}\left[ e^{-s_{1}P_{t,g}(s_{g,i}^2+h_{g}^2)^{-\frac{\eta_{g}}{2}}\Omega_{g,i}}\right]\right]\\
	&\stackrel{(b)}{=}\mathbb{E}_{\phi_g}\left[\prod_{s_{g,i}\in \phi_g \backslash b_0}^{}\frac{1}{1+s_{1}P_{t,g}\left(s_{g,i}^2+h_{g}^2\right)^{-\frac{\eta_{g}}{2}}}\right]\\
	&\stackrel{(c)}{=}\exp\left[-2\pi\lambda_g\int_{x_g}^{\infty}\left(1-\frac{1}{s_{1}P_{t,g}(z^2+x_g^2)^{-\frac{\eta_{g}}{2}}}\right)z\mathrm{d}z\right]
	\end{aligned}
\end{equation}
where (a) follows from the iid. distribution of the fading gain $\Omega_{g,i}$ and its independence of the point process $\phi_g$, (b) from the exponential distribution of $\Omega_{g,i}$ and (c) from the probability generation functional (PGFL) of the PPP of the interfering BSs locations~\cite{Abhishek} after replacing $s_{g,i}$ with $z$.	

To derive the Laplace transform of $I_a$, we denote by $u_{a,i}$ the distance from the $i$-th interfering UAV to the UE. The aggregate interference from all UAVs can be expressed as $\sum_{i=1}^{N}P_{t,a}u_{a,i}^{-\eta_{a}}\Omega_{a,i}$. Thus, its Laplace transform can be obtained as
	\begin{equation}\footnotesize
	\begin{aligned}
	\mathcal{L}_{I_a}(s_{1})&=\mathbb{E}_{I_a}\left[e^{-s_{1}I_a}\right]=\mathbb{E}_{I_a}\left[\exp\left(-s_{1}\sum_{i=1}^{N}P_{t,a}u_{a,i}^{-\eta_{a}}\Omega_{a,i}\right)\right]\\
	&\stackrel{(a)}{=}\mathbb{E}_{u_{a}}\left[\prod_{i=1}^{N} \mathbb{E}_{\Omega_{a}}\left(\exp\left(-s_{1}P_{t,a}u_{a,i}^{-\eta_{a}}\Omega_{a,i}\right)\right)\right]\\
	&\stackrel{(b)}{=}\mathbb{E}_{u_{a}}\left[\prod_{i=1}^{N} \left(1+\frac{s_{1}P_{t,a}u_{a,i}^{-\eta_{a}}}{m_{a}}\right)^{-m_{a}}\right]\\
	&\stackrel{(c)}{=}\left[\mathbb{E}_{u_{a}}\left[ \left(1+\frac{s_{2}P_{t,a}u_{a,i}^{-\eta_{a}}}{m_{a}}\right)^{-m_{a}}\right]\right]^{N}\\
	&\stackrel{(d)}{=}\left(\int_{E_{g}(x_{g})}^{w_{p}}f_{U}(u,E_g(x_g))\left(1+\frac{s_{1}P_{t,a}u^{-\eta_{a}}}{m_{a}}\right)^{-m_{a}}\mathrm{d}u\right)^{N}
	\end{aligned}
\end{equation}
where $f_{U}(u)$ is the distribution of the interferers distance from the reference UE given in Lemma 4 of~\cite{Chetlur} as $f_{U}(u,x)=\frac{f_W(u)}{\int_{x}^{w_p}f_W(w)\mathrm{d}w}$ where all the UAVs are further than $x$ from the reference UE. (a) follows from the iid distribution of the fading gains and from their independence of the interferers distance distributions. (b) follows from the moment generating functional (MGF) of the fading gain $\Omega_{a,i}$ that follows a gamma distribution. (c) follows from the iid distribution of the interferers distances. The final expression of $\mathcal{L}_{I_a}(s_{1})$ given in (8) is obtained by plugging $f_{U}(u,E_g(x_g))$ in the above equation. According to the association rule, when the UE associates to a BS, all the UAVs must be further than $E_{g}(x_g)$ where $x_{g}$ is the distance separating the reference UE and the serving BS. 
\subsection{Proof of Lemma  4}\label{app:distance}
The event $X_{g}>x_{g}$ is equivalent to the event of $R_{g}>x_g$ given that the reference UE associates with a BS, the CCDF of $X_g$ is given as
\begin{equation}
\bar{F}_{X_{g}}(x_g)=\mathbb{P}[R_{g}>x_g\mid n=g]=\frac{\mathbb{P}[R_g>x_g,n=g]}{\mathbb{P}[n=g]}
\end{equation}
where $R_g$ is the horizontal distance separating the nearest BS from the reference UE and $\mathbb{P}[n=g]=A_g$ is the BS association probability given in (2). The numerator of (32) is obtained as
\begin{equation}\small
\begin{aligned}
&\mathbb{P}[R_{g}>x_{g},n=g]\\
&=\mathbb{P}\left[R_{g}>x_{g},P_{t,g}(R_{g}^2+h_{g}^2)^{-\frac{\eta_{g}}{2}}>P_{t,a}R_{a}^{-\frac{\eta_{a}}{2}}\right]\\
&=\int_{x_{g}}^{\infty}\mathbb{P}\left[R_{a}>E_{g}(r)\right]f_{R_{g}}(r)\mathrm{d}r=\int_{x_{g}}^{\infty}\bar{F}_{R_a}\left(E_{g}(r)\right)f_{R_{g}}(r)\mathrm{d}r\\
\end{aligned}
\end{equation} 
where $E_{g}(r)$ and $\bar{F}_{R_a}(E_{g}(r))$ are given in (4) and (7), respectively. The CDF of $X_g$ is $F_{X_{g}}(x_g)=1-\bar{F}_{X_{g}}(x_g)$ and the PDF is given as
\begin{equation}\footnotesize
f_{X_{g}}(x_{g})=\frac{\mathrm{d}F_{X_{g}(x_{g})}}{\mathrm{d}x_{g}}=\frac{1}{A_{g}}f_{R_{g}}(x_{g})\left(\int_{E_{g}(x_{g})}^{w_{p}}f_{W}(w)\mathrm{d}w\right)^{N}
\end{equation}
where $f_{W}(w)$ is given in (5). Finally the distance distribution of $X_g$ can be obtained as in (8).
\subsection{Proof of Lemma 5}\label{app:LOS_association}
We start by providing the distributions of distances from the reference UAV to the nearest LOS BS in $\phi_L$ and the nearest NLOS BS in $\phi_N$. Denoting by $s_{L}$ the horizontal distance from the reference UAV to the nearest LOS BS in $\phi_L$, the CCDF of $s_{L}$ can be calculated as
\begin{equation}
\begin{aligned}
\bar{F}_{s_{L}}(s)&=	\mathbb{P}(s_{L} > s)= \mathbb{P}(\mathrm{No \; LOS \; BS \; closer \; than\;} s)\\
&=e^{-2\pi\lambda_g\int_{0}^{s}P_{\mathrm{LOS}}(r)r\mathrm{d}r}.
\end{aligned}
\end{equation}
Therefore, the CDF is $1-e^{-2\pi\lambda_g\int_{0}^{s}P_{\mathrm{LOS}}(r)r\mathrm{d}r}$ and the PDF can be found as
\begin{equation}
f_{s_{L}}(s)=	2\pi\lambda_g s P_{\mathrm{LOS}}(s) e^{-2\pi\lambda_g\int_{0}^{s}P_{\mathrm{LOS}}(r)r\mathrm{d}r}
\end{equation}
Similarly, the CDF and the PDF of the horizontal distance $s_{N}$ from the reference UAV to the nearest NLOS BS from $\phi_N$ are given as
\begin{equation}
\bar{F}_{s_{N}}(s)=e^{-2\pi\lambda_g\int_{0}^{s}(1-P_{\mathrm{LOS}}(r))r\mathrm{d}r}.
\end{equation}
and
\begin{equation}
f_{s_{N}}(s)=	2\pi\lambda_g s (1-P_{\mathrm{LOS}}(s)) e^{-2\pi\lambda_g\int_{0}^{s}(1-P_{\mathrm{LOS}}(r))r\mathrm{d}r}
\end{equation}
The reference UAV connects with a LOS BS in $\phi_L$ to get backhaul support if the nearest LOS BS has smaller path loss than that of the nearest NLOS BS in $\phi_N$. Thus, the LOS probability $A_\mathrm{L}$ that the reference UAV is associated with a LOS BS can be derived as follows
\begin{equation}\small
\begin{aligned}
A_\mathrm{L}&=\mathbb{P}\left[C_{L}z_{L}^{-\eta_{L}}>C_{N}z_{N}^{-\eta_{N}}\right]\\
&=\int_{0}^{\infty}\mathbb{P}\left[C_{L}(s_{L}^2+\Delta_h^2)^{-\frac{\eta_{L}}{2}}> C_{N}\left(s_{N}^2+\Delta_h^2\right)^{-\frac{\eta_{N}}{2}}\right]f_{s_{L}}(s_{L})\mathrm{d}s_{L}\\
&=\int_{0}^{\infty}\mathbb{P}\left[s_{N}>E_{L}(s_{L})\right]f_{s_{L}}(s_{L})\mathrm{d}s_{L}\\
&=\int_{0}^{\infty}\bar{F}_{s_{N}}(E_{L}(s_{L}))f_{s_{L}}(s_{L})\mathrm{d}s_{L}\\
\end{aligned}
\end{equation}
where 
\begin{equation}
\small
E_L(s_{L})=\sqrt{\left(\frac{C_{L}}{C_{N}}\right)^{\frac{2}{\eta_{N}}}({s_{L}}^2+\Delta_h^2)^{\frac{\eta_{L}}{\eta_{N}}}-\Delta_h^2}
\end{equation}
and $f_{s_{L}}(s)$ and $\bar{F}_{s_{N}}(s)$ are given in (36) and (37) respectively. By replacing $s_{L}$ with $x$, the LOS association probability $A_{L}$ is obtained as in (10).
\subsection{Proof of Lemma 6}\label{app:backhaul_distance}
Denote $X_L$ as the horizontal distance between the reference UAV and its serving LOS BS. Since the event $X_{L}>x$ is the event of $s_{L}>x$ given that the reference UAV connects to a LOS BS to get backhaul support, the probability of $X_{L}>x$ can be given as
\begin{equation}
\mathbb{P}[X_{L}>x]=\mathbb{P}[s_{L}>x\mid n=L]=\frac{\mathbb{P}[s_{L}>x,n=L]}{\mathbb{P}[n=L]}
\label{eq:cond_backhaul_distance_proof1}
\end{equation}
where $\mathbb{P}[n=L]=A_{L}$ is the probability that the UE associates to a LOS BS and follows from lemma 1. The joint probability of $s_{L}>x$ and $n=L$ is
\begin{equation}
\begin{aligned}
&\mathbb{P}[s_{L}>x,n=L]\\
&=\mathbb{P}\left[s_{L}>x,C_{L}(s_{L}^2+\Delta_h^2)^{-\frac{\eta_{L}}{2}}>C_{N}(s_{b,N}^2+\Delta_h^2)^{-\frac{\eta_{N}}{2}}\right]\\
&=\int_{x}^{\infty}\mathbb{P}[s_{N}>E_L(s_{L})]f_{s_{L}}(s_{L})\mathrm{d}s_{L}\\
&=\int_{x}^{\infty}\bar{F}_{s_{N}}(E_{L}(s_{L}))f_{s_{L}}(s_{L})\mathrm{d}s_{L}.
\end{aligned}
\label{eq:cond_backhaul_distance_proof2}
\end{equation}
Plugging (\ref{eq:cond_backhaul_distance_proof2}) in (\ref{eq:cond_backhaul_distance_proof1}) gives
\begin{equation}
\mathbb{P}[X_{L}>x]=\frac{1}{A_{L}}\int_{x}^{\infty}\bar{F}_{s_{N}}(E_{L}(s_{L}))f_{s_{L}}(s_{L})\mathrm{d}s_{L}.
\end{equation}
The CDF of $X_L$ is $F_{X_{L}}(x)=1-\mathbb{P}[X_{L}>x]$ and the PDF is given as 
\begin{equation}
f_{X_{L}}(x)=\frac{\mathrm{d}F_{X_{L}}(x)}{\mathrm{d}x}=\frac{1}{A_{L}}\bar{F}_{s_{N}}(E_{L}(x))f_{s_{L}}(x)\label{eq:cond_backhaul_distance_proof3}
\end{equation}
By plugging (36) and (37) in (\ref{eq:cond_backhaul_distance_proof3}), the PDF of the horizontal distance to the serving LOS BS is given as in (11) in Lemma~2.
Following the same procedure, the PDF of the horizontal distance between the reference UAV and its serving NLOS BS is determined and presented in (12).
\subsection{Proof of Theorem 1}\label{app:backhaul_probability}
Given that the reference UAV is connected to a BS in $\phi_L$, and that the desired signal link has a length of $s_{0}=x$, by Slivnyak’s Theorem, the conditional backhaul probability can be computed as
\begin{equation}\small
\begin{aligned}
&S_{L}(\tau_b)=\mathbb{P}[\mathrm{SINR}>\tau_b]\\
&=\int_{0}^{\infty}\mathbb{P}\left[\Omega_{b,0}>\frac{\tau_b(\sigma^2+I_b)}{P_{t,b}G_{0}C_{L}(x^2+\Delta_h^2)^{-\frac{\eta_{L}}{2}}}\right]f_{X_L}(x)\mathrm{d}x
\end{aligned}
\label{eq:theorem1_proof_1}
\end{equation}
where $I_b=I_{L}+I_{N}$ is the total interference from the LOS and the NLOS BSs, $\sigma^2$ is the noise power and $G_0$ is the maximum antennas gain. Noting that $\Omega_{b,0}$ is is a normalized gamma random variable with parameter $m_L$, we have the following approximation
\begin{equation}
\begin{aligned}
&\mathbb{P}\left[\Omega_{b,0}>\frac{\tau_b(\sigma^2+I_b)}{P_{t,b}G_{0}C_{L}(x^2+\Delta_h^2)^{-\frac{\eta_{L}}{2}}}\right]\\
&\stackrel{(a)}{\approx}\sum_{n=1}^{m_L}(-1)^{n+1}{m_\mathrm{L}\choose n}\mathbb{E}_{\phi_g}\left[e^{-\frac{n\gamma_L \tau_b (\sigma^2+I)(x^2+\Delta_h^2)^{\frac{\eta_{L}}{2}}}{P_{t,b}G_{0}C_{L}}}\right]\\
&\stackrel{(b)}{\approx}\sum_{n=1}^{m_L}(-1)^{n+1}{m_\mathrm{L}\choose n}e^{-n\mu_{L}\tau_b\sigma^2}\mathbb{E}_{\phi_g}\left[e^{-n\tau_b\mu_{L}I}\right]\\
&\stackrel{(c)}{\approx}\sum_{n=1}^{m_L}(-1)^{n+1}{m_\mathrm{L}\choose n}e^{-n\mu_{L}\tau_b\sigma^2}\mathcal{L}_{I_L}\left(n\mu_{L}\tau_b\right)\mathcal{L}_{I_N}\left(n\mu_{L}\tau_b\right)\\
\end{aligned}
\label{eq:theorem1_proof_2}
\end{equation} 
where (a) follows from~\cite{Alzer}, (b) from denoting $\mu_{L}=\frac{\gamma_L (x^2+\Delta_h^2)^{\frac{\eta_{L}}{2}}}{P_{t,b}G_{0}C_{L}}$, and (c) from denoting the Laplace functionals of the interference of the LOS BSs and the NLOS BSs as $\mathcal{L}_{I_{L}}(s)=\mathbb{E}[e^{-sI_{L}}]$ and  $\mathcal{L}_{I_{N}}(s)=\mathbb{E}[e^{-sI_{N}}]$, respectively, and the fact that $\phi_L$ and $\phi_N$ are independent.

Given that the desired backhaul link is LOS and has a length $x$, based on the minimum path loss association rule, all the LOS interfering BSs are farther than $x$, and all NLOS interfering BSs are farther than $E_L(x)$ from the reference UAV. The Laplace transform of the LOS interference $\mathcal{L}_{I_{L}}(t)$ where $t>0$, can be derived by applying the probability generating functional of a PPP~\cite{haenggi} as
\begin{equation}\footnotesize
\begin{aligned}
&\mathcal{L}_{I_{L}}(t)=\mathbb{E}\left[e^{-tI_{L}}\right]=\mathbb{E}\left[e^{-t\sum_{s_{b,i}\in \phi_g\backslash b_0}^{}P_{t,b}\Omega_{b,i}G_{b,i}C_{L}\left(s_{b,i}^2+\Delta_h^2\right)^{-\frac{\eta_L}{2}}}\right]\\
&=\exp\left(-2\pi\int_{x}^{\infty}\left(1-\mathbb{E}_{\Omega_{b},G_{b}}\left[e^{-tP_{t,b}\Omega_{b}G_{b}C_{L}\left(r^2+\Delta_h^2\right)^{-\frac{\eta_{L}}{2}}}\right]\right)\right.\\
&\times \lambda_g P_{\mathrm{LOS}}(r)r\mathrm{d}r\Bigg)
\end{aligned}\label{eq:theorem1_proof_3}
\end{equation}
Here, $\Omega_b$, $G_b$, and $r$ are dummy random variables for the small-scale fading, the gain and the distance in the interference channels. The term $\mathbb{E}_{\Omega_{b},G_{b}}\left[e^{-tP_{t,b}\Omega_{b}G_{b}C_{L}\left(r^2+\Delta_h^2\right)^{-\frac{\eta_{L}}{2}}}\right]$ in (\ref{eq:theorem1_proof_3}) can be computed as
\begin{equation}
\begin{aligned}
&\mathbb{E}_{\Omega_{b},G_{b}}\left[e^{-tP_{t,b}\Omega_{b}G_{b}C_{L}\left(r^2+\Delta_h^2\right)^{-\frac{\eta_{L}}{2}}}\right]\\
&\stackrel{(a)}{=}\sum_{k=1}^{4}p_{k}\mathbb{E}_{\Omega_{b}}\left[e^{-tP_{t,b}\Omega_{b}\bar{G}_{k}C_{L}\left(r^2+\Delta_h^2\right)^{-\frac{\eta_{L}}{2}}}\right]\\
&\stackrel{(b)}{=}\sum_{k=1}^{4}p_{k}\left(1+tP_{t,b}\bar{G}_{k}C_{L}\left(r^2+\Delta_h^2\right)^{-\frac{\eta_{L}}{2}}\right)^{-m_{L}}
\end{aligned}
\end{equation}
where (a) follows from the fact that the directivity gain in the interference channels $G_{b}$ is modeled as a discrete random variable, and (b) follows from computing the Laplace transform of the small-scale fading power $\Omega_{b}$ which follows a gamma distribution. Similarly, for the NLOS interfering links the Laplace transform $\mathcal{L}_{I_{N}}(t)$ is given as
\begin{equation}\footnotesize
\begin{aligned}
&\mathcal{L}_{I_{N}}(t)=\\
&\exp\left(-2\pi\int_{E_{L}(x)}^{\infty}\left(1-\mathbb{E}_{\Omega_{b},G_{b}}\left[e^{-tP_{t,b}\Omega_{b}G_{b}C_{N}\left(r^2+\Delta_h^2\right)^{-\frac{\eta_{N}}{2}}}\right]\right)\right.\\
&\times\lambda_g\left(1-P_{\mathrm{LOS}}(r)\right)r\mathrm{d}r\Bigg)
\end{aligned}
\label{eq:theorem1_proof_4}
\end{equation}
where
\begin{equation}
\begin{aligned}
&\mathbb{E}_{\Omega_{b},G_{b}}\left[e^{-tP_{t,b}\Omega_{b}G_{b}C_{N}\left(r^2+\Delta_h^2\right)^{-\frac{\eta_{N}}{2}}}\right]\\
&=\sum_{k=1}^{4}p_{k}\left(1+tP_{t,b}\bar{G}_{k}C_{N}\left(r^2+\Delta_h^2\right)^{-\frac{\eta_{N}}{2}}\right)^{-m_{N}}
\end{aligned}
\end{equation}
Finally, by plugging (\ref{eq:theorem1_proof_2}), (\ref{eq:theorem1_proof_3}) and (\ref{eq:theorem1_proof_4}) in (\ref{eq:theorem1_proof_1}) and by replacing $\mu_{L}$ by $\frac{\gamma_L \left(x^2+\Delta_h^2\right)^{\frac{\eta_{L}}{2}}}{P_{t,b}G_{0}C_{L}}$, we can get the expression in (14) of the conditional backhaul probability given that the UAV is served by a LOS BS. The same procedure can be followed to obtain the conditional backhaul probability given that the reference UAV is connected to a NLOS BS. Here, all NLOS interferers are farther than $x$ from the reference UAV and all LOS interferers are farther than $E_{N}(x)$. The detailed proof is omitted here and the expression of the NLOS backhaul probability $S_{\mathrm{N}}(\tau_b)$ is given in (15). Finally, by the law of total probability, the backhaul probability can be derived as in (13).
\subsection{Proof of Lemma 7}\label{app:pcova}
The conditional coverage probability $P_{cov,a}$ can be calculated as
\begin{equation}\footnotesize
\begin{aligned}
&P_{cov,a}=\mathbb{P}\left[\mathrm{SIR}\geq \beta | \mathrm{UAV},\mathrm{SINR}\geq\tau_b\right]\\
&=\mathbb{P}\left[\frac{P_{t,a}x_a^{-\eta_a}\Omega_a}{I_{agg,a}}\geq\beta | \mathrm{UAV},\mathrm{SINR}\geq\tau_b\right]\\
&\stackrel{(a)}{\approx}\mathbb{E}_{X_a}\left[\mathbb{E}_{I_{agg,a}}\left[\mathbb{P}\left(\Omega_{a}\geq\frac{\beta I_{agg,a}}{P_{t,a}x_{a}^{-\eta_a}}\right)\right]\right]\times S(\tau_b) \\
&\stackrel{(b)}{=}\mathbb{E}_{X_a}\left[\mathbb{E}_{I_{agg,a}}\left[\sum_{k=0}^{m_a-1}\frac{(\hat{I}_{a}+I_{g})^k}{k!}\times\left(\frac{m_a\beta x_a^{\eta_{a}}}{P_{t,a}}\right)^k\right.\right.\\
&\left.\left.\exp\left(-\left(\frac{m_a \beta x_{a}^{\eta_{a}}}{P_{t,a}}\right)\left(\hat{I}_a+I_{g}\right)\right)\right]\right] \times S(\tau_b)\\
&\stackrel{(c)}{=}S(\tau_b) \int_{h_a}^{w_p}\sum_{k=0}^{m_a-1}\frac{(-s_2)^k}{k!}\left[\frac{\partial^k}{\partial s_{2}^{k}}\mathcal{L}_{I_g}(s_2)\mathcal{L}_{\hat{I}_a}(s_2)\right]f_{X_a}(x_a)\mathrm{d}x_a
\end{aligned}
\end{equation}
where $s_2=\frac{m_a\beta x_a^{\eta_a}}{P_{t,a}}$ and $S(\tau_b)$ is the backhaul probability given in Theorem~$1$, (a) follows from the independence assumption of the two events, (b) is obtained from the CCDF of Nakagami-m fading UE-UAV channel power gain for $\Omega_{a}$ and from the expression of the aggregate interference $I_{agg,a}=\hat{I}_{a}+I_g$. Finally, (c) follows from the independence of $\hat{I}_a$ and $I_{g}$ and from the definition of the Laplace transform.
\bibliography{Arxiv_paper}
\bibliographystyle{ieeetr}
\end{document}